\documentclass[12pt,showkeys,showpacs] {revtex4}
\usepackage{graphicx}
\begin{document}

\title[Short Title]{Controlled Secure Direct Communication by Using GHZ Entangled State }

\author{Yan \surname{XIA}}
\author{Chang-Bao \surname{FU}}
\author{Feng-Yue \surname{LI}}
\author{Shou \surname{ZHANG}\footnote{E-mail: szhang@ybu.edu.cn}}
\affiliation{Department of Physics, College of Science, Yanbian
University, Yanji, Jilin 133002, PR China}

\author{Kyu-Hwang \surname{YEON}}

\affiliation{Department of Physics,  Institute for Basic  Science
Research, College of Natural Science, Chungbuk National University,
Cheonju, Chungbuk 361-763}

\author{Chung-In \surname{UM}}

\affiliation{Department of Physics, College of Science, Korea
University, Seoul 136-701, South Korea}

\begin{abstract} We present a controlled secure direct communication protocol
by using Greenberger-Horne-Zeilinger (GHZ) entangled state via
swapping quantum entanglement and local unitary operations. Since
messages transferred only by using local operations and a public
channel after entangled states are successfully distributed, this
protocol can protect the communication against a
destroying-travel-qubit-type attack. This scheme can also be
generalized to a multi-party control system.

 \end{abstract}

 \maketitle The quantum key distribution (QKD) is an ingenious
 application of quantum mechanics, in which two remote legitimate
 users (Alice and Bob) establish a shared secret key through the
 transmission of quantum signals. Since the pioneering work of Bennett and Brassard was published in
1984 \cite{BB84}, different quantum key distribution protocols
have been presented
\cite{ABCHPRA02,SJGPRA02,XLGPRA02,SPRA04,WPRL04}. Different from
the key distribution protocol, some quantum secure direct
communication (QSDC) protocols, which permit important messages to
be communicated directly without first establishing a random key
to encrypt them,  have been shown recently
\cite{BFPRL02,DLLPRA03,NBAPLA04,MZLCPL0518}.

Beige {\it et al.} \cite{ABCHPRA02} have proposed that messages
can be read out only after the transmission of an additional piece
of classical information for each qubit. Deng {\it et al.}
\cite{DLLPRA03} have proposed a two-step quantum direct
communication protocol using Einstein-Podolsky-Rosen (EPR) pair.
It was shown to be probably secure. Nguyen \cite{NBAPLA04} has
proposed an entanglement-based protocol for two people to
simultaneously exchange their messages. However, in all these
secure direct communication schemes, it is necessary to send the
qubits carrying secret messages in a public channel. Therefore,
Eve (eavesdropper) can attack the qubits in transmission.

Very recently, Man {\it et al.} \cite{MZLCPL0518} have proposed a
new deterministic secure direct communication protocol by using
swapping quantum entanglement
\cite{ZZHEPRL93,BVKPRA98,HSPRA00,PDGWZPRL01} and local unitary
operations. This communication protocol can be used to transmit
securely either a secret key or a plain text messages. In this
paper, we present a controlled secure direct communication by
using GHZ entangled state via swapping quantum entanglement and
local unitary operations. We show that the communication is secure
under some eavesdropping attacks \cite{ZMLPLA04,MZLCPL0522} and
that two users can transmit their secret messages securely.

Let us first describe the quantum entanglement swapping simply.
Let $|0\rangle$ and $|1\rangle$ be the horizontal and the vertical
polarization states of a photon, respectively. Then, the four Bell
states $|\psi^\pm\rangle=(|01\rangle \pm |10\rangle)/\sqrt{2}$ and
$|\phi^\pm\rangle=(|00\rangle \pm |11\rangle)/\sqrt{2}$ are
maximally entangled states in the two-photon Hilbert space. Let
the initial state of two-photon pairs be the product of any two of
the four Bell states, such as $|\psi^+_{12}\rangle$ and
$|\psi^+_{34}\rangle$. Then, after Bell measurements on the
photons 1 and 3 pair and the photons 2 and 4 pair, since the
equations

\begin{equation}\label{e1}
|\psi^+_{12}\rangle \otimes
|\psi^+_{34}\rangle=\frac{1}{2}(|\psi^+_{13}\rangle|\psi^+_{24}\rangle-|\psi^-_{13}\rangle|\psi^-_{24}\rangle
+|\phi^+_{13}\rangle|\phi^+_{24}\rangle-|\phi^-_{13}\rangle|\phi^-_{24}\rangle),
\end{equation}
\begin{equation}\label{e2}
|\psi^+_{12}\rangle \otimes
|\psi^-_{34}\rangle=\frac{1}{2}(|\psi^+_{13}\rangle|\psi^-_{24}\rangle-|\psi^-_{13}\rangle|\psi^+_{24}\rangle
-|\phi^+_{13}\rangle|\phi^-_{24}\rangle+|\phi^-_{13}\rangle|\phi^+_{24}\rangle),
\end{equation}
\begin{equation}\label{e3}
|\psi^+_{12}\rangle \otimes
|\phi^+_{34}\rangle=\frac{1}{2}(|\psi^+_{13}\rangle|\phi^+_{24}\rangle-|\psi^-_{13}\rangle|\phi^-_{24}\rangle
+|\phi^+_{13}\rangle|\psi^+_{24}\rangle-|\phi^-_{13}\rangle|\psi^-_{24}\rangle),
\end{equation}
\begin{equation}\label{e4}
|\psi^+_{12}\rangle \otimes
|\phi^-_{34}\rangle=\frac{1}{2}(|\psi^+_{13}\rangle|\phi^-_{24}\rangle-|\psi^-_{13}\rangle|\phi^+_{24}\rangle
-|\phi^+_{13}\rangle|\psi^-_{24}\rangle+|\phi^-_{13}\rangle|\psi^+_{24}\rangle),
\end{equation}
obviously hold, one can see that there is an explicit
correspondence between a known initial state of two qubit pairs
and its swapped measurement outcomes. In addition, it is easily
verified that the four Bell states can be transformed into each
other by some unitary operations. Then, encodings of secret
messages can be imposed on the Bell states by using the local
unitary operations, i.e.,
$\hat{I}|\psi^+_{34}\rangle=|\psi^+_{34}\rangle$,
$\hat{\sigma}_x|\psi^+_{34}\rangle=|\phi^+_{34}\rangle$,
$-i\hat{\sigma}_y|\psi^+_{34}\rangle=|\phi^-_{34}\rangle$, and
$\hat{\sigma}_z|\psi^+_{34}\rangle=|\psi^-_{34}\rangle$. We assume
that each of the above four unitary operations corresponds to a
two-bit encoding, i.e., $\hat{I}$ to 00, $\hat{\sigma}_x$ to 01,
$-i\hat{\sigma}_y$ to 10, and $\hat{\sigma}_z$ to 11. Taking
advantage of the quantum entanglement swapping and the assumption
of the two-bit codings, we can propose a controlled secure direct
communication by GHZ entangled state protocol.

Let us turn to depicting our controlled secure direct
communication protocol. First, we write the eight GHZ entangled
state basis in two different bases as follows:
\begin{eqnarray}\label{e5}
&|\psi_{1}\rangle&=\frac{1}{\sqrt{2}}(|000\rangle+|111\rangle)_{htc}\cr\cr&&=\frac{1}{2}[|+\rangle_h(|+\rangle_t|+\rangle_c+
|-\rangle_t|-\rangle_c)+|-\rangle_h(|+\rangle_t|-\rangle_c+|-\rangle_t|+\rangle_c)],
\end{eqnarray}
\begin{eqnarray}\label{e6}
&|\psi_{2}\rangle&=\frac{1}{\sqrt{2}}(|000\rangle-|111\rangle)_{htc}\cr\cr&&=\frac{1}{2}[|+\rangle_h(|-\rangle_t|+\rangle_c+
|+\rangle_t|-\rangle_c)+|-\rangle_h(|+\rangle_t|+\rangle_c+|-\rangle_t|-\rangle_c)],
\end{eqnarray}
\begin{eqnarray}\label{e7}
&|\psi_{3}\rangle&=\frac{1}{\sqrt{2}}(|100\rangle+|001\rangle)_{htc}\cr\cr&&=\frac{1}{2}[|+\rangle_h(|+\rangle_t|+\rangle_c+
|-\rangle_t|+\rangle_c)-|-\rangle_h(|+\rangle_t|-\rangle_c+|-\rangle_t|-\rangle_c)],
\end{eqnarray}
\begin{eqnarray}\label{e8}
&|\psi_{4}\rangle&=\frac{1}{\sqrt{2}}(|100\rangle-|001\rangle)_{htc}\cr\cr&&=\frac{1}{2}[|+\rangle_h(|+\rangle_t|-\rangle_c+
|-\rangle_t|-\rangle_c)-|-\rangle_h(|+\rangle_t|+\rangle_c+|-\rangle_t|+\rangle_c)],
\end{eqnarray}
\begin{eqnarray}\label{e9}
&|\psi_{5}\rangle&=\frac{1}{\sqrt{2}}(|010\rangle+|101\rangle)_{htc}\cr\cr&&=\frac{1}{2}[|+\rangle_h(|+\rangle_t|+\rangle_c-
|-\rangle_t|-\rangle_c)+|-\rangle_h(|+\rangle_t|-\rangle_c-|-\rangle_t|+\rangle_c)],
\end{eqnarray}
\begin{eqnarray}\label{e10}
&|\psi_{6}\rangle&=\frac{1}{\sqrt{2}}(|010\rangle-|101\rangle)_{htc}\cr\cr&&=\frac{1}{2}[|+\rangle_h(|+\rangle_t|-\rangle_c-
|-\rangle_t|+\rangle_c)+|-\rangle_h(|+\rangle_t|+\rangle_c-|-\rangle_t|-\rangle_c)],
\end{eqnarray}
\begin{eqnarray}\label{e11}
&|\psi_{7}\rangle&=\frac{1}{\sqrt{2}}(|110\rangle+|001\rangle)_{htc}\cr\cr&&=\frac{1}{2}[|+\rangle_h(|+\rangle_t|+\rangle_c-
|-\rangle_t|-\rangle_c)+|-\rangle_h(|-\rangle_t|+\rangle_c-|+\rangle_t|-\rangle_c)],
\end{eqnarray}
\begin{eqnarray}\label{e12}
&|\psi_{8}\rangle&=\frac{1}{\sqrt{2}}(|110\rangle-|001\rangle)_{htc}\cr\cr&&=\frac{1}{2}[|+\rangle_h(|+\rangle_t|-\rangle_c-
|-\rangle_t|+\rangle_c)+|-\rangle_h(|-\rangle_t|-\rangle_c-|+\rangle_t|+\rangle_c)],
\end{eqnarray}
where
\begin{equation}\label{e13}
|+\rangle=\frac{1}{\sqrt{2}}\ (|0\rangle+|1\rangle),
\end{equation}
\begin{equation}\label{e14}
|-\rangle=\frac{1}{\sqrt{2}}\ (|0\rangle-|1\rangle),
\end{equation}
{\it h} stands for `` {\it home}, '' {\it t} stands for `` {\it
travel}, '' and {\it c} stands for `` {\it control}. ''

(S1) Alice prepares a series ({\it N}) of GHZ entangled states in
\begin{eqnarray}\label{e15}
|\psi_1\rangle_{n}=\frac{1}{\sqrt{2}}(|000\rangle+|111\rangle)_{h_nt_nc_n},
\end{eqnarray}
where $n \in \{0,N\}$. She takes one photon from each GHZ
entangled state, say the photons ${\it t}_1, {\it t}_2, {\it t}_3,
{\it t}_4$, etc, and forms a string of photons in a regular
sequence, say the ordered string of photons is ${\it t}_1$${\it
t}_2$${\it t}_3$${\it t}_4\cdots$. She sends Bob the ordered
photon string
 (called the T sequence hereafter). In accordance with the ordering
 of the travel photons, Alice sends Charlie the photons ${\it c_n}$ (called the C sequence
 hereafter) and stores the photons ${\it h_n}$ (called the H sequence
 hereafter) by way of two photons as a group, i.e., photons ${\it
 h}_1$ and ${\it h}_2$ as group 1, photons ${\it h}_3$ and ${\it h}_4$ as group 2, etc.

 (S2) Bob and Charlie confirm that they have received the travel
 photons. Also, in a regular sequence, Bob stores the arrived
 photons by way of two photons as a group in terms of their
 order of arrival: that is, photons ${\it t}_1$ and ${\it t}_2$ as group 1,
 photons ${\it t}_3$ and ${\it t}_4$ as group 2, etc, and Charlie stores the arrived
 photons by way of two photons as a group in terms of their
 order of arrival: that is, photons ${\it c}_1$ an ${\it c}_2$ as group 1,
 photons ${\it c}_3$ and ${\it c}_4$ as group 2, etc.

 (S3) If Bob wants to transmit messages to Alice, he chooses
 randomly some photon groups as encoding-decoding groups (say, group 1 and 2,
 etc.) for his later two-bit encodings via local unitary
 operations {\it U} $(U_1=\hat{I}, U_2= \hat{\sigma}_x, U_3=-i\hat{\sigma}_y, U_4=\hat{\sigma}_z)$.
 The remaining photon groups are taken as checking groups. He
 publicly tells his choices to Alice and Charlie.

 (S4) For each photon in each checking group, Bob randomly
 chooses one of two sets of measuring basis (MB), $\{|0\rangle,
 |1\rangle\}$ or $\{|+\rangle, |-\rangle\}$, to measure his
 photons. Then, he publicly announces the order of the photons, the MB, and the measurement outcomes.
 Then, Alice and Charlie perform their measurements, under the same
 MB as that chosen by Bob, on the corresponding photons in their
 checking groups. The intercept-and-resend
attack, as well as the entangle-and-measure attack, can be
detected efficiently by this method. If their measurement outcomes
coincide when they use the same basis in the same order according
to
 Eq.~{(\ref{e5})$-$(\ref{e12})}, there are no Eves
 on the line. In this case, they continue to communication. Otherwise,
 they have to discard their transmission and abort the
 communication.

 (S5) Bob asks Charlie to carry out a Hadamard operation on each
 $c_n$ in order. The Hadamard operation has the form
 \begin{equation}\label{e16}
 H|0\rangle=\frac{1}{\sqrt{2}}(|0\rangle+|1\rangle),
 \end{equation}
  \begin{equation}\label{e17}
 H|1\rangle=\frac{1}{\sqrt{2}}(|0\rangle-|1\rangle),
 \end{equation}
 which will transform the state shown in Eq.~(\ref{e15}) into
 \begin{equation}\label{e18}
 |\psi_1\rangle_n=\frac{1}{2}[(|00\rangle+|11\rangle)_{h_n t_n} \otimes |0\rangle_{c_n}+(|00\rangle-|11\rangle)_{h_n t_n} \otimes
 |1\rangle_{c_n}].
 \end{equation}
Charlie measures photon ${\it c_n}$. If he obtains the outcome
$|0\rangle_{c_n}$, photons (${\it {h_n, t_n}}$) will collapse into
the state
\begin{equation}\label{e19}
|\eta_1^+\rangle_n=\frac{1}{2}(|00\rangle+|11\rangle)_{h_n t_n};
\end{equation}
otherwise, the state of photons (${\it {h_n, t_n}}$) will be
\begin{equation}\label{e20}
|\eta_1^-\rangle_n=\frac{1}{2}(|00\rangle-|11\rangle)_{h_n t_n}.
\end{equation}

(S6) After the above operations, in accord with the
encoding-decoding group ordering, Charlie informs Alice and Bob of
his measurement results via a classical communication.

(S7) In accord with the encoding-decoding groups ordering, Bob
performs his two-bit encodings via local unitary {\it U}
operations on the encoding-decoding groups according to his bit
strings (say, 0001$\cdots$) needing to be transmitted this time:
for instance, a $U_1$ operation on one photon of group 1 to
encoding 00, a $U_2$ operation on one photon of group 2 to encode
01, etc. After the unitary {\it U} operations, Bob performs the
Bell state measurements on all the encoding-decoding groups.

(S8) Bob publicly announces his Bell measurement result and the
encoding-decoding group's order for each encoding-decoding group.

(S9) Alice measures her unmeasured photon groups in Bell states
after Bob's public announcement in (S8). After she knows each of
Charlie and Bob's Bell measurement results with the group order
and her Bell measurement results with group orders, she can
identify the exact unitary {\it U} operations performed by Bob on
each encoding-decoding group. She can read the two-bit encodings.
For example, we suppose that Charlie's measurement results are
$|0\rangle_{c_1}$ and $|0\rangle_{c_2}$. Then, $|\psi_1\rangle_1$
and $|\psi_1\rangle_2$ are collapsed into the states
\begin{equation}\label{e21}
|\eta_1^+\rangle_1=\frac{1}{2}(|00\rangle+|11\rangle)_{h_1 t_1},
\end{equation}
\begin{equation}\label{e22}
|\eta_1^+\rangle_2=\frac{1}{2}(|00\rangle+|11\rangle)_{h_2 t_2}.
\end{equation}
 If Bob performs $U_1$ operation on one photon of group
1, then the state of the whole system composed of photons $({\it
h}_1$, ${\it t}_1$; ${\it h}_2$, ${\it t}_2)$ is
\begin{eqnarray}\label{e23}
U_1|\eta_1^+\rangle_1 \otimes
|\eta_1^+\rangle_2&=&\frac{1}{4}(|00\rangle+|11\rangle)_{h_1 t_1}
\otimes (|00\rangle+|11\rangle)_{h_2 t_2}\cr\cr&
=&\frac{1}{4\sqrt{2}}|\psi^+\rangle_{t_1
t_2}(|01\rangle+|10\rangle)_{h_1 h_2}+ \frac{1}{4\sqrt{2}}
|\psi^-\rangle_{t_1 t_2}(|01\rangle-|10\rangle)_{h_1
h_2}\cr\cr&&+\frac{1}{4\sqrt{2}}|\phi^+\rangle_{t_1
t_2}(|00\rangle+|11\rangle)_{h_1 h_2}+ \frac{1}{4\sqrt{2}}
|\phi^-\rangle_{t_1 t_2}(|00\rangle-|11\rangle)_{h_1
h_2}\cr\cr&=&\frac{1}{4}[|\psi^+\rangle_{t_1
t_2}|\psi^+\rangle_{h_1 h_2}+|\psi^-\rangle_{t_1
t_2}|\psi^-\rangle_{h_1 h_2}\cr\cr&&+|\phi^+\rangle_{t_1
t_2}|\phi^+\rangle_{h_1 h_2}+|\phi^-\rangle_{t_1
t_2}|\phi^-\rangle_{h_1 h_2}].
\end{eqnarray}
If Bob publicly announces his Bell measurement result is
$|\phi^+\rangle_{t_1 t_2}$ with a probability of 1/16, then
Alice's measurement result is $|\phi^+\rangle_{h_1 h_2}$. Thus,
she can conclude that Bob performed a $U_1$ operation on one
photon of group 1 and, therefore, extract the bits (00), and she
can read the two-bit encodings (see Table 1). Similarly, in
accordance with the photon group's ordering, she can obtain the
bit string (0001$\cdots$).
\begin{table}
\caption{Corresponding relations among the unitary {\it U}
operation (i.e., the encoding bits), the initial states, and Bob's
and Alice's Bell measurement results
($|\eta_1^\pm\rangle_n=|\eta_1^\pm\rangle_{h_n
t_n}=\frac{1}{2}(|00\rangle\pm|11\rangle)_{h_n t_n}$,
$|\mu_1^\pm\rangle_n=|\mu_1^\pm\rangle_{h_n
t_n}=\frac{1}{2}(|01\rangle\pm|10\rangle)_{h_n t_n}).$}\label{0}
\begin{tabular}{c|c|c|c}\hline
$U_1(00)$&$U_2(01)$&$U_3(10)$&$U_4(11)$\\
\hline $|\eta_1^+\rangle_{h_1 t_1} \otimes |\eta_1^+\rangle_{h_2
t_2}$&$|\mu_1^+\rangle_{h_1 t_1} \otimes |\eta_1^+\rangle_{h_2
t_2}$&$i|\mu_1^-\rangle_{h_1 t_1} \otimes |\eta_1^+\rangle_{h_2
t_2}$ &$|\eta_1^-\rangle_{h_1 t_1} \otimes |\eta_1^+\rangle_{h_2
t_2}$
\\
\hline\hline$|\phi^{+}\rangle_{t_1 t_2},|\phi^+\rangle_{h_1
h_2}$&$|\phi^{+}\rangle_{t_1 t_2},|\psi^+\rangle_{h_1
h_2}$&$|\phi^{+}\rangle_{t_1 t_2},|\psi^-\rangle_{h_1
h_2}$&$|\phi^{+}\rangle_{t_1 t_2},|\phi^-\rangle_{h_1 h_2}$
\\
\hline$|\phi^{-}\rangle_{t_1 t_2},|\phi^-\rangle_{h_1
h_2}$&$|\phi^{-}\rangle_{t_1 t_2},|\psi^-\rangle_{h_1
h_2}$&$|\phi^{-}\rangle_{t_1 t_2},|\psi^+\rangle_{h_1
h_2}$&$|\phi^{-}\rangle_{t_1 t_2},|\phi^+\rangle_{h_1 h_2}$
\\
\hline$|\psi^{+}\rangle_{t_1 t_2},|\psi^+\rangle_{h_1
h_2}$&$|\psi^{+}\rangle_{t_1 t_2},|\phi^+\rangle_{h_1
h_2}$&$|\psi^{+}\rangle_{t_1 t_2},|\phi^-\rangle_{h_1
h_2}$&$|\psi^{+}\rangle_{t_1 t_2},|\psi^-\rangle_{h_1 h_2}$
\\
\hline$|\psi^{-}\rangle_{t_1 t_2},|\psi^-\rangle_{h_1
h_2}$&$|\psi^{-}\rangle_{t_1 t_2},|\phi^-\rangle_{h_1
h_2}$&$|\psi^{-}\rangle_{t_1 t_2},|\phi^+\rangle_{h_1
h_2}$&$|\psi^{-}\rangle_{t_1 t_2},|\psi^+\rangle_{h_1 h_2}$
\\
 \hline
\end{tabular}
\end{table}

(S10) Similarly, if Charlie wants secure direct communication with
Alice, he can do the same thing as Bob does.

(S11) The controlled secure direct communication has been
successfully completed.

In the present protocol, we use the property of quantum
entanglement swapping instead of the property of quantum
entanglement. If the sender (Bob) wants to transmit messages to
the receiver (Alice), Charlie must take an Hadamard operation on
each of his photons, measure it, and tell the results to Bob and
Alice in order. When GHZ entangled state is successfully shared,
no qubit has to be exchanged in a quantum channel. This leads to
four important advantages. Firstly, If no eavesdropping is found
in the checking procedure, the secret messages can be transmitted
successfully. Because there is not a transmission of the qubit
that carries the secret messages between Alice and Bob in a public
channel, it is completely secure for controlled secure direct
communication if a perfect quantum channel is used. That is, after
the checking procedure, there is no longer any chance for Eve to
attack the secret messages. Secondly, the degradation of the
entanglement of the photon pair will not decrease due to the
travel of the photon in the H sequence in a real quantum channel.
Thirdly, after insuring the security of the quantum channels, if
Charlie would like to help Alice and Bob to communicate, Alice and
Bob can communicate secret messages directly under the control of
the third side Charlie. Only with the help of controller Charlie,
the sender and the receiver can implement secure direct
communication successfully. Fourthly, this protocol can also be
generalized to a multi-party control system in which {\it N}
parties share a large number of {\it N}-particle GHZ entangled
states. Party {\it M} ($M \in N$) as receiver, party {\it Q} ($Q
\in N, Q \neq M$) as sender, and $N-2$ parties, except for party
{\it M} and party {\it Q}, as controllers. The multi-party
controlled secure direct communication can also succeed.

In summary, we have proposed a controlled secure direct
communication protocol by using a large number of  GHZ entangled
states via entanglement swapping and local unitary operations.
When GHZ entangled state is successful shared,  no qubit has to be
exchanged in a quantum channel. Different form Man's scheme
\cite{MZLCPL0518}, in this protocol, anyone of the multi-partners
can send messages to any receiver secretly with the help of
controllers by using a local operation and a reliable public
channel. The result shows that for such a protocol, can realize
successful controlled secure direct communication between two
users. Since the message transferred only by using local
operations and public channels after entanglement was successfully
distributed, this protocol can protect the communication against
the destroying-travel-qubit-type attack. This scheme can also be
generalized to a multi-party control system.

\begin{center}
{\bf ACKNOWLEDGMENTS}
\end{center}

This work was supported by the Korea Science and Engineering
Foundation and by the National Natural Science Foundation of China
under Grant No 60261002.

\end{document}